\documentclass[twoside,10pt,a4paper]{newFNLstyle}
\usepackage[]{graphicx}
\usepackage{cite}

\begin{document}

\volnumpagesyear{0}{0}{000--000}{2001}
\dates{received date}{revised date}{accepted date}

\title{Long-range fluctuations of random potential landscape 
as a mechanism of $1/f$ noise in 
hydrogenated amorphous silicon}

\authorsone{Boris V. Fine\thanks{Present address of B. V. F.: 
Physics Department, University of Tennessee, 1413 Circle Dr., Knoxville, TN 37996, USA. E.mail:bfine@utk.edu}}
\affiliationone{Max Planck Institute for the Physics of Complex Systems,
Noethnitzer Str. 38, D-01187 Dresden, Germany \\
and  \\
Spinoza Institute, P.O. Box 80195, 3508 TD Utrecht, The Netherlands}

\authorstwo{ Jeroen P. R. Bakker and Jaap I. Dijkhuis}
\affiliationtwo{Debye Institute,
Utrecht University, P.O. Box 80000, 3508 TA Utrecht, 
The Netherlands}

%\authorinfo{E.mail of B. V. F.: fine@mpipks-dresden.mpg.de}
\maketitle

\markboth{Long-range fluctuations of random potential landscape ...}{Fine, Bakker and Dijkhuis}

\pagestyle{myheadings}
% Comment this out to remove the running heads

% Keywords have to before the abstract I'm afraid.

\begin{abstract}
We describe a mechanism, which links  the long-range 
potential fluctuations induced by charged defects to the low frequency
resistance noise widely known as $1/f$ noise.  This mechanism  
is amenable to the first principles microscopic calculation of 
the noise spectrum, which includes the absolute noise intensity.
We have performed such a calculation for the thin films of
hydrogenated amorphous silicon (a-Si:H) under the condition that current 
flows perpendicular to the plane of the films, and found a very
good agreement between the theoretical noise intensity and the measured one.
The mechanism described is 
quite general. 
It should be present in a broad class of systems 
containing poorly screened charged defects.
\end{abstract}

%%%%%%%%%%%%%%%%%%%%%%%%%%%%%%%%%%%%%%%%%%%%%%%%%%%%%%%%%%%%%

\section{Introduction}
\label{introduction}
In this work we present a theoretical and experimental study 
of low-frequency
voltage noise in $\mu$m-thick films of amorphous silicon (a-Si:H) 
under the condition that electric current 
flows perpendicularly to the plane of the 
films. This phenomenon is associated with resistance fluctuations, which,
in the presence of current, manifest themselves as voltage noise. 
The spectrum of this noise is close to $1/f$, where $f$ is the frequency.

Our motivation for this work is two-fold. On the one hand, 
we describe a new microscopic mechanism of $1/f$ noise, 
which should be present in a variety of systems.
On the other hand, the particular material studied, a-Si:H, 
is very important technologically because of its
applications in various 
photo-voltaic devices (such as e.g. solar cells and thin-film transistors). 
The 
technological importance of a-Si:H represents the further advantage, that
this material has been heavily studied in the past,
and, therefore, its microscopic characteristics, which we need as 
input to our theory, are reasonably well known\cite{Street}.

Among many existing proposals aimed at the general description of $1/f$ 
noise,
perhaps, the most successful one is  the model of an 
ensemble of two-state systems
having a broad distribution of activation energies 
(BDAE)\cite{vanderZiel,duPre,DDH}. 
The BDAE model gives a reasonable explanation for the generic nature of
spectral shapes close to $1/f$  and, furthermore, 
(after Dutta, Dimon and Horn\cite{DDH}) 
predicts the temperature dependence
of the small deviations from that shape.  It is the experimental observation
of those small deviations in a variety of systems\cite{Weissman}
that constitutes the strongest evidence for the adequacy of the BDAE model.
However, this model as such does not represent
a full theoretical description of 
the noise, because it does not address the origin 
of the two-state systems and the mechanism by which they 
couple to the resistance fluctuations. As a result, the absolute 
intensity of the $1/f$ noise
remains an adjustable parameter. 

In principle, it is not obvious at all that a universal 
mechanism should underlie every occurrence of $1/f$ noise in the systems, 
that seem to obey the BDAE description.
However, the unsatisfactory reality is that there are no examples 
(at least, we are not aware of any), when  a microscopic mechanism of 
the $1/f$ noise
has been worked out  in full detail for one ``BDAE'' system, and,
at the same time, 
the results of the
calculation based on that mechanism have agreed with experiment.
Here  ``full detail''
means: (i) the identification of microscopic fluctuators with their activation 
energies and activation rates; (ii) identification of the mechanism that
couples those fluctuators to the resistance noise; and (iii) 
a first principles
calculation of the absolute noise intensity.  Such a ``full detail''
treatment was recently given by us for a-Si:H in Ref.~\cite{FBD}~. 
This
treatment, however, included many material-specific details, 
not all of which were crucially important for the 
understanding the noise mechanism.
In this work, we describe the same experimental setting as 
in Ref.~\cite{FBD}
but do so in a more intuitive way by replacing some detailed calculations with
simple estimates. It turns out that the outcome of such a description is
not much different from that of the full calculation.

\section{Film characteristics and experimental details.}
\label{experiment}

We study an $n-i-n$ 
film of \mbox{a-Si:H}, where $n$ denotes an  
electron 
doped layer, and $i$  an  undoped layer. 
The thickness of each of the
$n$-layers is 40~nm, while the thickness of the $i$-layer is 
\mbox{$d = 0.91\mu$m}. 
The area of the film is $A=0.56\hbox{cm}^2$. 
The film was grown by plasma enhanced chemical vapor deposition (PECVD). 
It was subsequently thermally annealed and, afterwards, kept protected from light.
The setup of our noise experiments was the same as in Ref.~\cite{Verleg}. 

We observed voltage noise spectra 
at frequencies $f = 1 \div 10^4\,$Hz and temperatures
$T = 340 \div 434\,$K  in the presence of electric current 
flowing perpendicularly to the plane of the film. The film itself was
thus sandwiched between highly conductive contact layers. 
The voltages applied
were small enough to correspond to the linear part of the
film's I-V characteristic.
The experiments were performed without illumination of the film.

\section{Previous studies of $1/f$ noise in a-Si:H}
\label{overview}

The previous studies of  $1/f$ noise in the films of 
a-Si:H\cite{BA,BAV,PK,PIK,KK,Verleg,VD,GJK,Goennenwein,Johanson} 
illustrate
the general  situation described in Section~\ref{introduction}: 
 although a large body of diverse 
phenomenological information about the noise 
has been collected, the problem of identifying
the origin of the noise has remained open.

In this work 
we focus only on one out of many possible situations that have been studied 
experimentally: perpendicular current, no illumination, 
undoped main resistivity layer, annealed sample. In that  case,
it was shown experimentally\cite{Verleg} that (i) the statistics of
noise is Gaussian and (ii) the temperature dependence of 
the spectral slope (parametrizing the small deviations from the exact $1/f$
dependence) agrees well with the BDAE model.

It was evident
since the early studies\cite{BA}, that the timescales 
of activation for the electron escape from deep defects in a-Si:H 
correspond  well
to the frequency range where the noise was observed. The problem was 
that, even though the energies of the defect levels have a relatively 
broad distribution, the Fermi factor limits the noise to the levels, 
which are
located in the thermal window around the chemical potential\cite{BAV}. 
This thermal 
window is not broad enough to give a spectral shape close to $1/f$.

%In other words, in the absence of the theory allowing one to compute the 
%absolute noise intensity, the discussions were focused on the origin 
%of the broad distribution of the activation energies, and, if 
%such a distribution could not be identified in the context of a 
%given mechanism, the mechanism itself was ruled out. 

If, nevertheless, one insists on computing the noise intensity associated 
with the fluctuations of the defect occupation numbers, then 
there are still two general scenarios: the resistance 
fluctuations  can be due to  the fluctuations of either the number of free 
carriers 
or their mobility. The fluctuations of the number of free carriers can be
produced by the random emission and capture 
of free electrons by deep defects. This possibility was investigated by 
Verleg and Dijkhuis\cite{Verleg}  (albeit on the basis of a very simple
model). They came to conclusion that the noise intensity
due to this mechanism would be several orders of magnitude smaller than the 
one observed, and, furthermore, the timescale of such a noise would be 
controlled by the fast capture times rather than the slow emission times,
which shifts the characteristic noise frequency away 
from the observation range. Later, a more detailed theoretical 
investigation of this type of noise has supported the 
conclusion of a low noise intensity\cite{Bakker}.

A simple realization of mobility fluctuations could be related to the 
change of the defect cross-sections. 
This scenario, however, has the problem that the relative 
concentration of deep defects is too small to affect the transport 
of free carriers. The mobility of free carriers is 
controlled mainly by the elastic scattering from the short-range 
inhomogeneities of the amorphous lattice structure.

Other mechanisms of the $1/f$ noise in a-Si:H
have also been proposed, including thermally activated chemical processes\cite{BAV},
generation-recombination processes  with various degrees of 
sophistication\cite{Verleg,VD,BCFD},
and, finally,  resistance networks near percolation threshold\cite{LK} 
(in the context of coplanar currents). 

Although, in principle, each of these proposals can be viable, 
none of them has been tested conclusively so far.

\section{Outline of the noise mechanism proposed in this work}
\label{outline}

In the following, we shall revive the old idea that the noise is caused
by the fluctuations of the defect occupation numbers. That idea, however,
is complemented by a novel noise mechanism, which involves 
long-range potential fluctuations induced by 
fluctuations of the defect charges. These potential fluctuations 
then cause the 
fluctuations of the local densities of conduction electrons, 
which, in turn, lead to the observed resistance fluctuations.

It is important to realize that 
the above mechanism should certainly be 
present in the material studied. Its theoretical 
description is quite straightforward and, at the same time, gives the value of 
integrated noise intensity without adjustable parameters. Therefore,
if such a theory produces the noise intensity comparable 
with the one observed in experiment, then it is quite unlikely that another 
noise mechanisms contributes to the experimental spectrum on the top of the one 
just described.

Our theoretical description also contains a new, though very simple, idea 
that the distribution of the  fluctuation rates of defect charges 
comes not just from the distribution of
the energy levels of electrons bound to the defects, 
but also from the distribution of the activation barriers , which electrons 
have to overcome to escape from the defects.
Unlike the former distribution, the later one is not truncated by the 
Fermi factor\cite{BAV,Weissman} and thus can underlie the spectral 
shape close to $1/f$.

In the rest of this paper, our theory is exposed in 
Sections~\ref{q-of-interest}-\ref{evaluation} followed
by comparison with experiments (Section~\ref{comparison})
and conclusions(Section\ref{conclusions}). Central to our treatment is
Section~\ref{relation}, which contains the description of the 
noise mechanism.

\section{Quantity of interest}
\label{q-of-interest}

The voltage noise spectrum can be expressed as:
\begin{equation}
{S_V(f) \over V^2} = 4 \int_0^\infty C_V(t) \hbox{cos}(2 \pi f t) \ dt,
\label{Sv}
\end{equation}
where $V$ is the applied voltage, and 
\begin{equation}
C_V (t) = { \langle \delta V(t) \delta V(0) \rangle \over V^2} =
{ \langle \delta R(t) \delta R(0) \rangle \over R_0^2}.
\label{Cv1}
\end{equation}
Here, 
$\delta V(t)$ is the voltage fluctuation, 
$R_0$ is the average resistance of the film, and $\delta R(t)$ is the 
equilibrium resistance fluctuation. 

The second equality in 
Eq.(\ref{Cv1}) follows from the assumption of constant current $I$ 
flowing through 
the film, i.e. $\delta V(t) = I \delta R(t)$ and $V = I R_0$.
The assumption of constant current is granted, because 
(i) the current noise of
external origin is suppressed by a very large 
resistance connected in series with the film; and 
(ii) the current noise of "internal" origin manifests itself at
the equilibrium Johnson-Nyquist noise\cite{Johnson,Nyquist}, 
which was measured independently
with zero applied voltage and then subtracted from the spectrum
taken with non-zero voltage. 

The link between the resistance noise
and the voltage noise has also been established experimentally by 
observing that  $S_V(f)$ defined as the difference between the spectra
at zero applied voltage and non-zero applied 
voltage is proportional to $V^2$. 

\section{Description of the resistivity layer}
\label{band}

In the following, we consider a somewhat idealized problem of resistance
noise coming from a resistivity layer of thickness $z_r$ and 
volume $V_r = z_r A$ having uniform
material characteristics.  Because of the band bending, 
most of the resistance of our actual 
film originates from the center of the intrinsic layer. 
The effective thickness of that central layer 
is\cite{FBD} $z_r = 0.26\ \mu$m. 

The density of states of undoped a-Si:H  
is characterized by a band gap of 
$1.8\,$~eV 
between the mobility edges $E_v$ and $E_c$
in the valence and conduction bands, respectively. 
The defect states, which play an important role in the noise mechanism,
are located deep inside the band gap. Because of the proximity of the 
n-layers, the chemical potential 
$\mu$ is significantly closer to $E_c$ than to $E_v$. (From the measurements 
of the conductivity activation energy, we estimate that $E_c - \mu = 0.63$~eV.)
As a result, the number of electrons in the conduction band 
is much greater than the number of holes in the valence band,
i.e. conduction electrons are the primary carriers of electric current.

One parameter, which is particularly 
important for the rest of this work, is $r_s$, the screening radius of 
deep defects.  
In order to estimate it,
we first note that the conventional mechanism of screening by 
conduction electrons is not operational in our film, because,
at the temperatures of experiment,  their concentration 
($10^{10} - 10^{13} \hbox{cm}^{-3}$) is much smaller than
the concentration of deep defects 
($\sim 10^{16} \hbox{cm}^{-3}$.
Instead, we identify two screening mechanisms: (i) by contact layers
and (ii) by other deep defects. The effective screening radius due to
both
screening mechanisms was estimated in Ref.\cite{FBD} as 0.2~$\mu$m.

In order to simplify the
theoretical description, we shall assume that $r_s \ll z_r $,
i.e., in this sense, we consider
three-dimensional ``bulk'' limit. 
In our  film, $r_s \sim z_r $.
However, the noise intensity computed with the actual values of $r_s$ and 
$z_r$ differs  from the outcome of the bulk limit calculation 
only by factor of two\cite{FBD}.  

We also assume that the film is
still thin enough, so that an electron emitted 
from a deep defect is much more likely 
to escape into the contact layers than to be captured
by another deep defect. This means that the charges of
different deep defects fluctuate independently.
Such an assumption is well applicable to the experiments 
with transverse currents\cite{FBD},
because, in these experiments, 
the contact layers spread over the entire film surface, which means
that any defect in the resistivity layer is no more than half of 
the film thickness away from the contacts.
However, the same assumption 
is not applicable to the experiments with co-planar 
currents\cite{PK,PIK,KK,GJK,Johanson,Kasap,BK}, 
where the distance to contacts is of the order of the in-plane dimensions
of the films.

\section{Relation between the resistance fluctuations and 
the long-range fluctuations of the local potential}
\label{relation}

Now we describe the fluctuations of the resistivity  within the
  resistivity layer. These fluctuations arise
as a consequence of the  fluctuations  of 
the screened Coulomb potential $\phi(t, {\mathbf{r}})$
created by deep charged defects, known as dangling bonds:
\begin{equation}
\phi(t, {\mathbf{r}}) = \sum_i {\Delta q_i(t) \over 
\epsilon |{\mathbf{r}} - {\mathbf{a}}_i|}
\hbox{exp} \left( -{|{\mathbf{r}} - {\mathbf{a}}_i| \over r_s }  \right).
\label{phi}
\end{equation}
Here 
$\Delta q_i(t)$ is the fluctuation of the $i$th defect charge
with respect
to its average value, ${\mathbf{a}}_i$  the position   the 
defect, $r_s$ the screening radius, and
\mbox{$\epsilon =  12$} the dielectric constant.
The defects involved 
may be located outside of the resistivity layer.

When the  potential $\phi(t, {\mathbf{r}})$ fluctuates, 
the mobility  edge
tracks it, i.e.
\begin{equation}
E_c(t, {\mathbf{r}}) = E_{c0} + e \phi(t, {\mathbf{r}}),
\label{Ect}
\end{equation}
where  $e$ is the electron charge.
Since the chemical potential $\mu$ does 
not shift with $e \phi(t, {\mathbf{r}})$, 
the density of conduction electrons, $n_e$, re-equilibrates  
following $E_c(t, {\mathbf{r}})$ on the timescale of electron drift
from the center of the $i$-layer to the $n$-layers. 
Because of the strong band bending inside the $i$-layer\cite{FBD},
the drift takes less than  $10^{-7}$s,
i.e. the re-equilibration is effectively instantaneous on the 
timescales of the noise studied \mbox{(${2 \pi \over f} \sim 10^{-4}\div 1$~s).} 
The fluctuating quasi-equilibrium density 
of the conduction electrons
is then  proportional to 
$\exp \left[-(E_c(t, {\mathbf{r}}) - \mu)/ k_B T \right]$,
where $k_B$ is 
the Boltzmann constant.  Finally, the fluctuating local resistivity $\rho$,
which is  
is
inversely proportional to $n_e$, can be written as
\begin{equation}
\rho(t, {\mathbf{r}}) = 
X \exp \left({E_c(t, {\mathbf{r}}) - \mu \over k_B T} \right),
\label{rhotr}
\end{equation}
where $X$ is a proportionality coefficient. 

Assuming for a moment [and proving later] that 
\begin{equation}
|e \phi(t, {\mathbf{r}})| \ll k_B T,
\label{ephiT}
\end{equation} 
 we expand
\begin{equation}
\rho(t, {\mathbf{r}}) = \rho_0 + \delta \rho(t, {\mathbf{r}}),
\label{rhotr1}
\end{equation}
where 
\begin{equation}
\rho_0 = X \hbox{exp} \left( {E_{c0}  - \mu \over k_B T} \right),
\label{rho01}
\end{equation}
and
\begin{equation}
\delta \rho(t, {\mathbf{r}}) = 
{e \phi(t, {\mathbf{r}}) \over k_B T} \rho_0.
\label{drho1}
\end{equation}
For $\delta \rho \ll \rho_0$, the fluctuation of the total resistance
(derived in the Appendix) is
\begin{equation}
\delta R(t) =  {1 \over A^2} 
\int_{\cal{V}} \delta \rho(t, {\mathbf{r}})  d^3 r,
\label{dR}
\end{equation}
where $\cal{V}$ is the space inside
the resistivity layer (limited by $\pm z_r/2$ along the $z$-axis
and by the edges of the film in the $xy$-plane).
Substituting $R_0 = z_r \rho_0/A$ and $\delta R(t)$ given by Eq.(\ref{dR}) 
into Eq.(\ref{Cv1}), 
and then using
Eq.(\ref{drho1}), we obtain
\begin{equation}
C_V(t) =  
\left({e
\over
k_B T V_r}\right)^2
\int_{\cal{V}} d^3 {\mathbf{r}} \int_{\cal{V}}  d^3 {\mathbf{r}}^{\prime}  
\langle \phi(t, {\mathbf{r}})  \phi(0, {\mathbf{r}}^{\prime}) \rangle.
\label{Cvphi}
\end{equation}
Given Eq.(\ref{phi}), the correlation function of potential fluctuations can be written as
\begin{equation}
\langle \phi(t, {\mathbf{r}})  \phi(0, {\mathbf{r}}^{\prime}) \rangle =
\sum_{i,j} {\langle \Delta q_i(t) \  \Delta q_j(0) \rangle \over 
\epsilon^2 |{\mathbf{r}} - {\mathbf{a}}_i |  \  | {\mathbf{r}}^{\prime} - {\mathbf{a}}_j  |} \
\hbox{exp} \left[ - { |{\mathbf{r}} - {\mathbf{a}}_i | +  | {\mathbf{r}}^{\prime} - {\mathbf{a}}_j  | \over r_s } \right]
\label{phiphitr}
\end{equation}

Equation (\ref{Cvphi}) 
should have a very broad range of applicability. All the material-specific details
affect only the evaluation of $\langle \phi(t, {\mathbf{r}})  \phi(0, {\mathbf{r}}^{\prime}) \rangle$. 
In Ref.\cite{FBD}, we have performed this evaluation taking into account numerous microscopic characteristics
of a-Si:H. Below, however, we present a cruder estimate, which is more intuitive
and yet reasonably accurate.

The time dependence of $\langle \phi(t, {\mathbf{r}})  \phi(0, {\mathbf{r}}^{\prime}) \rangle$
in Eq.(\ref{phiphitr}) comes from the correlators of charge fluctuations 
$\langle \Delta q_i(t) \  \Delta q_j(0) \rangle$. For $i \neq j$, 
$\langle \Delta q_i(t) \  \Delta q_j(0) \rangle =0$, 
because, as discussed in Section~\ref{band}, the charge fluctuations of different defects
are independent of each other.  The non-zero contribution comes from 
the correlators
with $i=j$, which can be expressed as
\begin{equation}
\langle \Delta q_i(t) \  \Delta q_i(0) \rangle = 
\langle \Delta q_i^2 \rangle \hbox{exp} \left(-{t \over \tau_i} \right).
\label{dqidqi}
\end{equation}
Here $1/\tau_i$ is the fluctuation rate of the $i$th defect. The correlator
is characterized by a single exponent, because every defect is assumed to have only
two states neutral or charged (with charge either $+e$ or $-e$). 
The potential fluctuations
are induced mainly by the defects, which we call ``thermally active" or simply ``active."
These active defects have binding energies $E$ in the thermal window $\pm 2 k_B T$
around the chemical potential $\mu$. Their concentration is, therefore,
\begin{equation}
n_T = 4 \ k_B T \ D(\mu), 
\label{nT}
\end{equation}
where $D(\mu)$ is the density of defect states around the chemical potential. It can be
estimated as 
\begin{equation}
D(\mu) = {n_D \over 2 \Delta E}, 
\label{D}
\end{equation}
where $n_D$ is the total concentration of deep defects, and
$\Delta E$ is the half-width of the distribution of their binding energies.
We have found that in our film\cite{FBD} 
$n_D \approx 6 \ 10^{15} \ \hbox{cm}^{-3}$,
and $\Delta E \approx 0.15 $~eV.

In comparison with the rest of the defects, the thermally active ones have the 
largest amplitude of charge fluctuations. The absolute value of their charge 
has roughly the same probability to be 0 or $e$. Therefore, its mean value is
$e/2$, and the mean squared amplitude of fluctuations is
\begin{equation}
\langle \Delta q_T^2 \rangle \approx {e^2 \over 4}.
\label{qT}
\end{equation}

Now we estimate $\langle \phi(t, {\mathbf{r}})  \phi(0, {\mathbf{r}}^{\prime}) \rangle$ by 
taking the average over the spatial distribution of active defects (assumed to be random)
and over the distribution of their relaxation times (inverse fluctuation rates) $P_{\tau}(\tau)$.
For this estimate we use the following {\it Ansatz}:
\begin{equation}
\langle \phi(t, {\mathbf{r}})  \phi(0, {\mathbf{r}}^{\prime}) \rangle =
\langle \phi^2 \rangle \ 
\hbox{exp} \left[ - { |{\mathbf{r}} - {\mathbf{r}}^{\prime} | \over r_s } \right] \ 
\int \hbox{exp} \left(-{t \over \tau} \right) \ P_{\tau}(\tau) \ d \tau,
\label{phiphi0}
\end{equation}
where, from Eqs.(\ref{phiphitr},\ref{nT},\ref{qT}),
\begin{equation}
\langle \phi^2 \rangle \equiv \langle \phi(0, 0)^2 \rangle =
\int {\langle \Delta q_T^2 \rangle \over \epsilon^2 \ {r^{\prime \prime}}^2 } 
\ \hbox{exp} \left(-{2 r^{\prime \prime} \over r_s} \right) \ 
n_T \ 4 \pi^2 {r^{\prime \prime}}^2 \  dr^{\prime \prime} = 
{2 \pi e^2 r_s D(\mu) k_B T \over \epsilon^2},
\label{phi2}
\end{equation}
where $r^{\prime \prime}$ is an integration variable corresponding to
$|{\mathbf{r}} - {\mathbf{a}}_i|$ in Eq.(\ref{phiphitr}).
From Eq.(\ref{phi2}), the value of $|e \phi(t, {\mathbf{r}})|$ can 
be estimated as:
$e \sqrt{\langle \phi^2 \rangle}  \sim 3.5\hbox{meV}$.
Since $k_B T \sim 30$meV, 
the  assumption (\ref{ephiT}) was adequate.

Substituting Eqs.(\ref{phiphi0},\ref{phi2}) into Eq.(\ref{Cvphi}) 
and integrating over 
${\mathbf{r}}$ and ${\mathbf{r}}^{\prime}$ 
(under the assumption $r_s \ll z_r$ made in Section~\ref{band}),
we obtain
\begin{equation}
C_V(t) = {16 \pi^2 e^4 r_s^4 D(\mu) \over \epsilon^2 k_B T V_r} \ 
\int \hbox{exp} \left(-{t \over \tau} \right) \ P_{\tau}(\tau) \  \ d \tau .
\label{Cvtau}
\end{equation}

Now we discuss the origin of 
the distribution of the activation times $P_{\tau}(\tau)$.

\section{Activation barriers}
\label{barriers}

\begin{figure}
\setlength{\unitlength}{0.1cm}
%=======================================================================
\begin{picture}(50, 58)
{
\put(0, 0){
\includegraphics[height=2.4in]{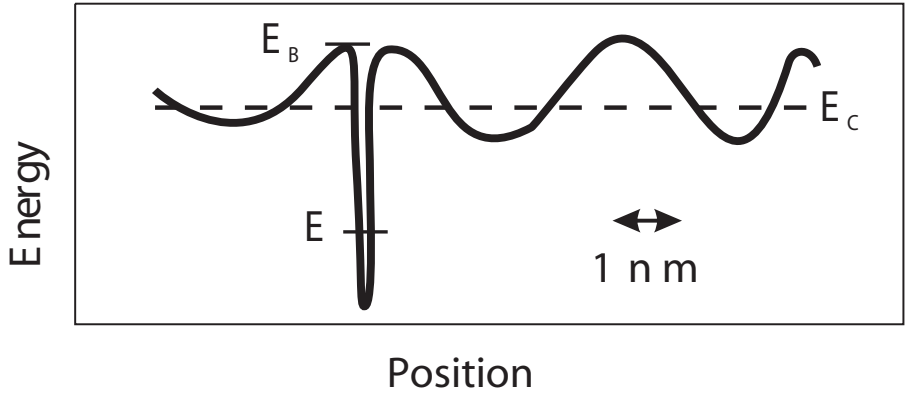}
}
}
\end{picture} 
%============== 
\caption{Cartoon of a deep defects surrounded by medium-range structural
disorder. Note: $e \phi(t, {\mathbf{r}})$ fluctuates on a much longer
length scale and with much smaller amplitude.} 
\label{fig1} 
\end{figure}

In order to escape from a deep defect, an electron should reach 
the mobility edge $E_c$.  However, the activation barriers $E_B$
(indicated in Fig.\ref{fig1}) can vary
as a
result of the {\it medium-range} disorder of 
the amorphous structure (on a length scale of $1 \div 10\,$nm). 
The activation time of 
a thermally active defect should then read
\begin{equation}
\tau(E_B) = \tau_0 \ \hbox{exp} \left(  {E_B - \mu \over k_B T} \right),
\label{tau}
\end{equation}
where $\tau_0$ is the prefactor of the order of $10^{-13}$~s \cite{FBD}.

In Ref.\cite{FBD}, we have  assumed Gaussian probability
distribution for the values of $E_B$:
\begin{equation}
P(E_B) =  
{ 1 \over \sqrt{2 \pi} \Delta E_B} 
\exp \left( -{(E_B - E_{B0})^2 \over 2  \Delta E_B^2} \right).
\label{PEB}
\end{equation}
where $E_{B0}$ and $\Delta E_B$ were 
extracted from the experimental spectra. These were the only
two adjustable parameters in our treatment. 
We have found that $E_{B0} - \mu = 0.9$~eV (0.27 eV above $E_{c0}$), 
while $\Delta E_B = 0.09$~eV

Such an adjustment, however, does not compromise the experimental 
tests of the theory. 
The values of $E_{B0}$ and $\Delta E_B$ affect the spectral 
shape but not the 
integrated noise intensity. 
Therefore, if the theory predicts a too small
noise prefactor, the adjustment of these two parameters only 
redistributes the spectral intensity in the range of observations 
but cannot make the theoretical spectra agree with the experimental ones.
In the next Section, we shall proceed with an estimate of
the prefactor in front of the (approximate) $1/f$ spectral dependence, 
which is independent of the assumption of the Gaussian shape of $P(E_B)$,
but instead relies only on the crude value of $\Delta E_B \sim 0.1$~eV.
Since $\Delta E_B$ is a characteristics
of the energy landscape in a-Si:H, one can hardly expect 
that it has a much different value. We shall also estimate the
integrated noise intensity, which does not depend on the value of
$\Delta E_B$ at all.

Although the idea that
 the activation 
barriers should be distributed is very simple, it has not been 
exploited previously.
One issue here is whether the 
barriers are long enough or high enough to ensure that the activation 
processes dominate the tunneling under the potential landscape. 
Given that
not much is reliably known about the random 
potential landscape on the scale of $1 \div 10\,$nm, it is only
obvious that the relatively high temperatures of experiment favor the activation
processes. From our crude theoretical estimates, 
the  activation over a barrier, which is 0.3~eV high and 3.5~nm, 
long starts dominating
the tunneling under the same barrier at $T=340 K$. 
For smaller barrier heights or higher temperatures,
the critical length of the barrier becomes smaller.

However, our main argument in favor of the existence 
of the distribution of activation 
barriers is purely empirical:

The mechanism we describe remains perfectly valid, if one assumes
that the activation barriers are not distributed at all, i.e. 
there exists only one activation barrier for all defects: $E_B = E_{c0}$,
and, therefore,
$P(E_B) = \delta(E_B - E_{c0})$. In that case,
the theoretical spectral intensity predicted by our treatment
would significantly 
exceed the experimental one in the higher frequency part
of the observed spectrum at $T=340 K$. 
One would then have to explain, how another 
noise mechanism {\it suppresses} the spectral intensity due to the 
present one --- a task, which  seems to be extremely difficult if not
impossible.

\section{Evaluation of the noise spectrum}
\label{evaluation}

Substituting Eq.(\ref{Cvtau}) into Eq.(\ref{Sv}) and also using Eq.(\ref{tau}) to switch from 
integration over 
$\tau$ to integration over $E_B$, we obtain
\begin{equation}
{S_V(f) \over V^2} = {64 \pi^2 e^4 r_s^4 D(\mu) \over \epsilon^2 k_B T V_r}
\int_{- \infty}^{+ \infty} 
{\tau(E_B) P(E_B) dE_B \over 1 + 4 \pi^2 f^2 \tau^2(E_B)}
\label{Sv1}
\end{equation}

Since
the proximity to the $1/f$ spectral shape results only from the fact that 
the distribution $P(E_B)$ is much broader than $k_B T$, we obtain the prefactor
in front of the $1/f$ dependence by substituting the constant value
\begin{equation}
P(E_B) = {1 \over 2 \Delta E_B}
\label{PEB1}
\end{equation}
into Eq.(\ref{Sv1}), which gives
\begin{equation}
{S_V(f) \over V^2} \approx 
{8 \pi^2 e^4 r_s^4 D(\mu) \over \epsilon^2  V_r \Delta E_B}.
\ {1 \over f}
\label{Sv2}
\end{equation}
The above estimate cannot be applicable to all frequencies,
because Eq.(\ref{PEB1}) explicitly violates
the normalization condition. Nevertheless, expression (\ref{Sv2}) 
constitutes a good approximation in a broad frequency domain around
the frequency corresponding to the 
maximum of the probability distribution 
$P(E_B)$.  

One can also obtain the integrated noise intensity (of course,
not from the approximation (\ref{Sv2}) but from Eq.(\ref{Sv1})): 
\begin{equation}
\int_0^{\infty} {S_V(f) \over V^2} df  \equiv C_V(0) = 
{16 \pi^2 e^4 r_s^4 D(\mu) \over \epsilon^2 k_B T V_r}.
\label{Svint}
\end{equation}
If the estimate (\ref{D}) for $D(\mu)$ is substituted into 
Eqs.(\ref{Sv1},\ref{Sv2}), then the ``bulk limit''
results obtained from a more accurate description and reported 
in Ref.\cite{FBD} can be recovered.

The remarkable fact about expressions (\ref{Sv2},\ref{Svint}) is that,
even though the noise mechanism rests
on the fluctuations of the number of conduction electrons,
the resulting spectrum  is 
independent of their equilibrium concentration. 
Furthermore, the distinct feature of
Eq.(\ref{Sv2}) is that the noise prefactor does not depend on temperature. 
This should be contrasted with the popular empirical law due to 
Hooge\cite{Hooge},
according to which the noise prefactor is inversely proportional to the
number of (thermally activated) carriers and thus
decreases exponentially as temperature increases.
As far as the defect characteristics are concerned, then the
noise intensity depends only on one of them, namely, $D(\mu)$, the density 
of the
defect states at the chemical potential.
This dependence is, in fact, weaker than the simple proportionality to
$D(\mu)$, and may even exhibit the opposite trend, 
because deep defects in undoped a-Si:H screen each other, and therefore, 
the screening radius $r_s$ decreases
with the increase of $D(\mu)$ (see Ref.\cite{FBD}).

\section{Comparison with experiment}
\label{comparison}

\begin{figure}
\setlength{\unitlength}{0.1cm}
%=======================================================================
\begin{picture}(50, 85)
{
\put(-30, -213){
\includegraphics[height=13in]{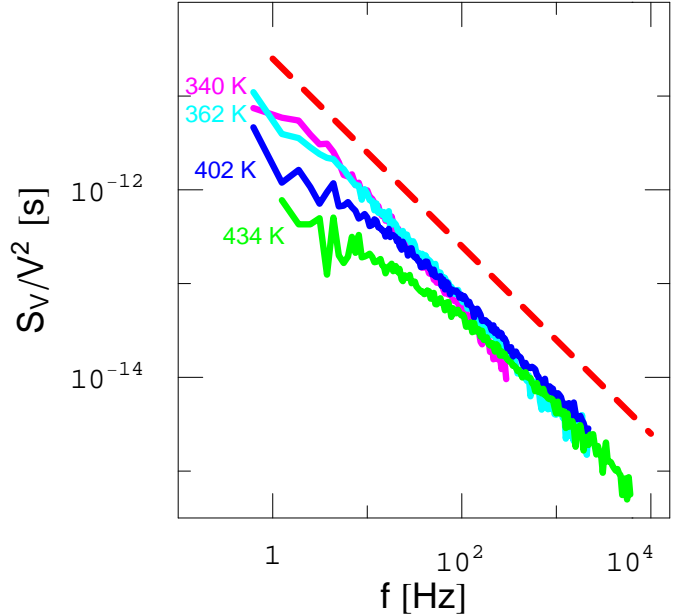} 
 }
}

\end{picture} 
%============== 
\caption{(a) Noise spectra: 
solid lines represent the experimental data taken at four different temperatures
indicated in the plot; dashed line represents the theoretical  
prediction (\protect{\ref{Sv2}}). 
} 
\label{fig2} 
\end{figure}

In Fig.~\ref{fig2}, the theoretical spectrum (\ref{Sv2}) 
is compared with the experimental ones taken at four different temperatures.
The experimental spectra are the same as reported in Ref.~\cite{FBD}. They were 
obtained by subtracting the zero-current noise from the total noise observed
with $V=50$~meV. The numbers substituted into the theoretical 
spectrum (\ref{Sv2}) are the following: $r_s = 0.2 \ \mu$m, 
$\Delta E_B = 0.1 \ $eV,
$\epsilon = 12$, $V_r = z_r A = 0.26 \ \mu\hbox{m} \times 0.56 \ \hbox{cm}^2$, 
$D(\mu)$ is obtained from the estimate (\ref{D}) with 
$n_D = 6 \ 10^{15} \hbox{cm}^{-3}$ and $\Delta E = 0.15 $~eV.

One can observe that 
(i) the experimental spectra  at four different temperatures
strongly overlap with each other, in agreement with the temperature independent
form of the theoretical expression (\ref{Sv2}); and (ii) the absolute value
of the experimental spectra agree within factor of three with the value given
by Eq.(\ref{Sv2}). 

\begin{figure}
\setlength{\unitlength}{0.1cm}
%=======================================================================
\begin{picture}(50, 60)
{
\put(-25, -197){
\includegraphics[height=11in]{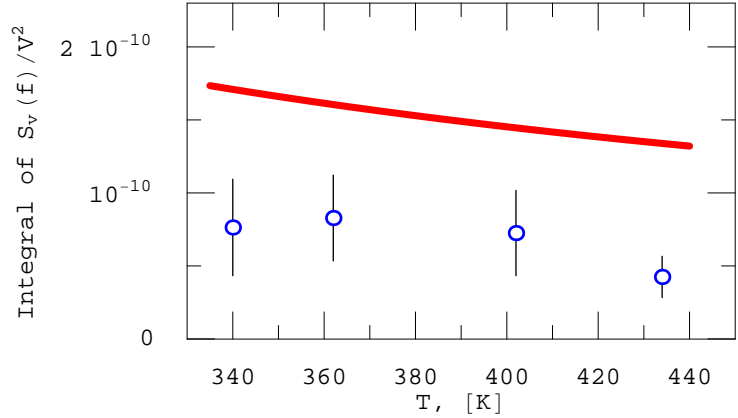}
 }
}
\end{picture} 
%============== 
\caption{Integrated noise intensity. Empty dots
represent the experimental values for the four spectra shown in
Fig.~\ref{fig2}. The error bars 
on the experimental points are obtained as described in the text.
The solid line corresponds to the theoretical expression (\ref{Svint}).  
} 
\label{fig3} 
\end{figure}

In Fig.~\ref{fig3}, we present the comparison between the theoretical 
(Eq.(\ref{Svint}))
and the experimental values of the integrated noise intensity (obtained
with the numbers given above).  
Since the window of experimental observation does not extend over the infinite 
range of frequencies,
we employed the following extrapolation procedure, which 
entailed large but quantifiable uncertainties.
First, we obtained the lower ends of the error bars by integrating the
experimental noise
spectra only in the frequency range of the actual experimental observations.
Then, the upper ends  were obtained
by making the power law extrapolations of the spectra 
beyond the frequency range of observation 
(up to $10^{-6}$~Hz for small frequencies and $10^8$~Hz for large
frequencies), and then
adding the integrals over the extrapolated tails to the lower end values
of the error bars.
Finally, the ``experimental''
points indicated in Fig.~\ref{fig3} were  chosen as
the middle points of the above error bars.

Given the relatively crude estimates, which were involved at
various stages of the derivation of Eqs.(\ref{Sv2},\ref{Svint}), 
and the uncertainty
of the values of $r_s$, $D(\mu)$ and $\Delta E_B$, the agreement is, in fact,
very good. In particular, since
$r_s$
enters Eqs.(\ref{Sv2},\ref{Svint}) in the fourth power,
the uncertainty in the value of $r_s$ is the single largest source
of error in the theoretical predictions. 
The factor of three discrepancy would vanish if,
e.g. $0.15 \ \mu$m  were used for $r_s$ instead of $0.2\ \mu$m.
In the present case, however, most of the discrepancy
can be attributed not to
the uncertainty in the value of $r_s$ but to a controllable theoretical error.
Namely, the application of the ``bulk limit'' $r_s \ll z_r$ to the situation, where 
$r_s \sim z_r$, increases the value of the theoretical noise intensity
by factor of two\cite{FBD}. It has also been shown in
Ref.~\cite{FBD} that the calculation {\it a la} 
Dutta, Dimon and Horn\cite{DDH},
which uses the
Gaussian distribution of energy barriers (Eq.(\ref{PEB})), 
can account 
for the temperature-dependent suppression
of the noise intensity at the low-frequency end of 
the experimental spectra. 

Finally, we should mention that  two additional experimental 
tests of the present theory, which involve 
different sandwich structures, 
are reported by us in an different work\cite{BCFD1}.

\section{Conclusions}
\label{conclusions}

In conclusion, we have described a microscopic mechanism of 
$1/f$ noise in $n-i-n$ sandwich structures of a-Si:H.  
A very good agreement between this description and our experiments
clearly indicates that in the 
frequency domain $1-10^4$~Hz, the noise mechanism proposed is responsible 
for at least 
a substantial fraction of the noise intensity observed in experiment. 
Since this mechanism 
is quite general, its applicability to a broader class of materials
merits further investigation.

%%%%%%%%%%%%%%%%%%%%%%%%%%%%%%%%%%%%%%%%%%%%%%%%%%%%%%%%%%%%%
\section*{ACKNOWLEDGMENTS}       
 
We thank R. E. I. Schropp for providing us with the sample. 
The work of \mbox{B. V. F.} was supported in part by the 
Netherlands Foundation ``Fundameteel Onderzoek der Materie'' (FOM)
and
the ``Nederlandse Organisatie voor Wetenschapelijk Onderzoek'' (NWO).

%%%%%%%%%%%%%%%%%%%%%%%%%%%%%%%%%%%%%%%%%%%%%%%%%%%%%%%%%%%%%

\section*{APPENDIX}

Here we derive the relationship (\ref{dR}) 
between small resistivity fluctuations and the resistance fluctuations.

We shall assume that the spatial coarse-graining,
which in the following underlies continuous integration,
is of the order of 0.01$\mu$m. It is thus
much greater than the ballistic mean free path $l_b \approx 5 \div 10$\AA.
Therefore, the electric current flowing through each coarse-grained element
can be characterized by Ohm's law:
\begin{equation}
{{\mathbf{E}}} ({\mathbf{r}}) = \rho({\mathbf{r}}) {\mathbf{j}} ({\mathbf{r}}),
\label{Ohm}
\end{equation}
where ${{\mathbf{E}}}$ is the electric field, $\rho$ the resistivity,
${\mathbf{j}}$ the current density, and ${\mathbf{r}} \equiv (x,y,z)$ 
the position in the sample. The $z$-axis is chosen along the direction 
of total current, i.e. perpendicular to the plane of
the film.

The potential difference across the resistivity layer 
is given by integral
\begin{equation}
V = \int_{-z_r/2}^{z_r/2} E_z(x_0, y_0,z) dz = 
\int_{-z_r/2}^{z_r/2} \rho(x_0, y_0,z)  j_z(x_0, y_0,z) dz,
\label{VE}
\end{equation}
where $x_0$ and $y_0$ are just two arbitrary coordinates in 
the plane of the film. 
In the following, however, we shall use a somewhat redundant 
but equivalent expression:
\begin{equation}
V = {1 \over A} \int_{-z_r/2}^{z_r/2} dz \int_{\cal{A}} dx \ dy \
\rho(x,y,z) \ j_z(x, y, z) ,
\label{Vint}
\end{equation}
which represents the average over the equal values of 
the voltage difference over the area $\cal{A}$ of the film.

As a zero approximation, we consider a layer spreading along the 
$z$-axis from $-z_r/2$ to $z_r/2$ and having uniform resistivity $\rho_0$.
We thus represent the total resistivity as 
\begin{equation}
\rho(t, {\mathbf{r}}) = \rho_0 + \delta \rho(t, {\mathbf{r}}),
\label{rho}
\end{equation}
where $\delta \rho(t, {\mathbf{r}})$ is a small correction caused
by the long-range fluctuations of the local potential 

In general, the resistivity fluctuations are accompanied
by the fluctuations $\delta {\mathbf{j}} (t, {\mathbf{r}})$ 
of the  current density. 
The expression for the total current density is thus 
\begin{equation}
{\mathbf{j}} (t, {\mathbf{r}}) = {\mathbf{j}}_0 + \delta {\mathbf{j}} (t, {\mathbf{r}}),
\label{j}
\end{equation}
where
\begin{equation}
{\mathbf{j}}_0 = \left( 0, 0, {I \over A} \right).
\label{j0}
\end{equation}
The fact, that the total current through the film should stay 
constant in the presence of the resistance fluctuations, imposes the
following constraint:
\begin{equation}
\int_{\cal{A}}  \  \delta j_z(t, x, y, z_0) \ dx \ dy = 0,
\label{djint}
\end{equation}
where $z_0$ is an arbitrary coordinate between $-z_r/2$ and $z_r/2$.

Linearizing Eq.(\ref{Vint}) with respect to $\delta \rho$ and
$\delta {\mathbf{j}}$ we obtain
\begin{equation}
\delta V(t) = {1 \over A}  
\int_{\cal{V}} \left[ 
\delta \rho(t, {\mathbf{r}}) j_{0z}  +
\rho_0 \delta j_z(t, {\mathbf{r}})
\right] \  dx \ dy \ dz ,
\label{dVt}
\end{equation}
where $\cal{V}$ refers to the three-dimensional space limited by $\pm z_r/2$
along the $z$ axis and by the edges of the film in the $xy$-plane.
The integration of the second term in Eq.(\ref{dVt}) gives zero by virtue
of constraint (\ref{djint}). 
Thus, recalling that $j_{0z} = I/A$, we obtain Eq.(\ref{dR})
\begin{equation}
\nonumber
\delta R(t) = {\delta V(t) \over I} = {1 \over A^2} 
\int_{\cal{V}} \delta \rho(t, {\mathbf{r}}) \  d^3 r.
\label{dR1}
\end{equation}

\bibliographystyle{spiebib_Fine}   %>>>> makes bibtex use spiebib.bst

\end{document}